\documentstyle[12pt, fleqn]{article}
\renewcommand{\baselinestretch}{1.3}
\newcommand{\be}{\begin{equation}}
\newcommand{\ee}{\end{equation}}
\newcommand{\bea}{\begin{eqnarray}}
\newcommand{\eea}{\end{eqnarray}}
\newcommand{\su}{\mbox{$SU_q(2)$\,\,}}
\newcommand{\pro}{\partial}
\newcommand{\op}{\omega^{++}}
\newcommand{\om}{\omega^{--}}
\newcommand{\oo}{{\omega^0}}
\newcommand{\dfrac}{\displaystyle\frac}
\newcommand{\ba}{\begin{array}}
\newcommand{\ea}{\end{array}}
\newcommand{\dpst}{\displaystyle}
\newcommand{\dpp}{{\cal D}^{++}}
\newcommand{\dmm}{{\cal D}^{--}}
\newcommand{\docal}{{\cal D}^{0}}

\newcommand{\vareps}{\varepsilon}
\newcommand{\rmm}{({\hat R}^{-1})}
\newcommand{\qin}{\displaystyle\frac{1}{q}}
\newcommand{\uone}{\mbox{$U(1)$\,\,}}

\newcommand{\nn}{\nonumber}
\newcommand{\pbar}{\bar{\phi}}
\begin{document}
\title{Three Dimensional Differential Calculus on the Quantum Group \su and Minimal Gauge Theory}
\author{ D.G. Pak \thanks {E-mail: dmipak@apctp.kaist.ac.kr}}
\date{}
\maketitle
\begin{center}
       Department of Theoretical Physics, Research Institute\\
       of Applied Physics, Tashkent State  University,\\
       Vuzgorodok, 700095, Tashkent, Republic of Uzbekistan \\
\end{center}

\begin{center}
Classification numbers: 210, 220, 240
\end{center}

\begin{abstract}
     Three-dimensional bicovariant differential calculus on the quantum
 group $SU_q(2)$  is constructed using the approach based on global covariance under the
action of the
stabilizing subgroup $U(1)$. Explicit representations of possible
$q$-deformed Lie algebras are obtained in terms of differential operators.
The consistent gauge covariant differential calculus on \su is uniquely
defined. A non-standard Leibnitz rule is proposed
 for the exterior differential. Minimal gauge theory with \su
quantum group symmetry is considered.
\end{abstract}
 \vskip 50pt

 \newpage
\section{Introduction}
\indent
%\stepcounter{section}
One of the features of non-commutative geometry in the
quantum group theory [1-5]
 is non-uniqueness
in defining a differential calculus on the quantum groups and quantum spaces.
The bicovariance condition determines a unique differential calculus on the
linear quantum groups $GL_{q}(N)\,$ (up to symmetry corresponding to the
exchange $q \rightarrow \qin$) [6, 7]
 and
provides existence of the corresponding
 gauge covariant differential algebra [8].
Direct reducing the $GL_{q}(N)$-bicovariant differential calculus to
a case of the special linear quantum group $SL_{q}(N)\,$
encounters  difficulties connected with a loss
 of the centrality condition for
a quantum determinant. Four-dimensional $4D_{\pm}$ \, bicovariant and
three-dimensional $(3D)$ left-covariant differential calculi on the
simplest special unitary
quantum group \su were considered as well using a
standard Woronowicz approach [3, 6].
 A full consistent construction of the 3D bicovariant differential calculus
 and a gauge covariant differential algebra on the \su
are unknown up to now, furthermore, there are strong
limitations imposed by no-go theorems [8].
 A possible way to solve this
 problem suggests using a non-standard Leibnitz rule as it was
considered in ref. [9].

    In this paper possible 3D bicovariant differential calculi and
gauge covariant differential algebra on the quantum group \su
are considered in the framework of approach which respects a global
 $U(1)$-covariance.
 The group \uone is a stabilizing subgroup for the quantum
 group \su and the $U(1)$-covariant treatment allows to pass
straightforward to the description of the quantum sphere
 $S_{q}^{2} \, \sim \, {\su } / {\uone}$. In Section 2 we
 construct
explicit representations of $q$-deformed Lie algebras of left-invariant
vector fields on \su in terms of differential operators.
The $U(1)$-covariance constraint reduces the variety of possible
covariant
differential calculi on \su and leads to a unique gauge covariant
differential algebra as it is shown in Section 3.
We propose non-standard Leibnitz rules for the exterior differential
which are compatible with the quantum group \su structure and gauge
covariance.
Section 4 is devoted to construction of the minimal quantum group
gauge theory of \su.

\section{ $q$-deformed Lie algebras}
\setcounter{equation}{0}
%\stepcounter{section}
Following the $R$-matrix formalism [4] the main commutation
relation for the generators $T^i_j \;\;
(i, j =1,2)$ of the quantum group \su is defined by a
standard $R$-matrix as follows
\be
   R_{12}T_1T_2 = T_2T_1R_{12}.        \label {mainrel}
\ee
Let us choose a covariant parametrization for the matrix
 $T^i_j$
\bea
 T^i_j=\left(\begin{array}{cc}
                  y^1&x^1                              \label {param}    \\
                  y^2&x^2
                 \end{array} \right)
                             \equiv (y^ix^i),
\eea
where  $x^i, y_i \  $ are generators (coordinates)
of the function algebra on the quantum hermitean
vector space $U^2_q$  endowed with an involution
${\ast}:{ \stackrel{\ast}{x^{i}} = y_i}$
and $SU_q(2)$-comodule structure.The unimodularity condition
                 takes a simple covariant form
\bea
\ba{cc}
  {\cal D} \equiv {\det}_{q} T^i_j = x_i y^i = 1,&
                     x_i=\vareps_{ij}x^j.                       \nn
\ea
                                                                \nn
\eea
Hereafter the \su indices are raised and lowered with the
invariant metric
$\vareps_{ij} (\vareps_{12}=1, \vareps_{21}=-\frac{1}{q})$.

     The parametrization (\ref{param}) was used in a harmonic
formalism [10] of extended superfield supesymmetric theories. The
coordinates $(x, y)$ parametrize the
quantum sphere
 $S^{2}_q  \, \sim \,  {\su} / {\uone}$
and are just the quantum generalizations of classical
 harmonic functions $(u^{\pm})$ (so called "harmonics")
 \bea
     x^i \equiv u^{+i}, \,\,\,\,\,\,\,y^i\equiv u^{-i}. \nn
\eea
The signs $(\pm)$ correspond to  charges $(\pm1)$
of a stabilizing subgroup \uone for the quantum
group \su. To simplify notations we shall not pass to the
notations adopted in the harmonic formalism keeping in mind
that all geometric objects (like coordinates, derivatives, differential
forms
 etc.) have definite \uone charges.

    Consider main commutation relations between the coordinates
$(x,y)$ and derivatives
 $\partial_i\equiv \dfrac{\partial}{\partial
 x^i}, \;\;
 \bar\partial^i\equiv \dfrac{\partial}{\partial y_i}$ on the
quantum group \su:
 \bea
\ba{cc}
R_{12}(\pro_T)_1 (\pro_T)_2 = (\pro_T)_2 (\pro_T)_1 R_{21} ,&
    (\pro_T)^i_j \equiv \left(\begin{array}{cc}
                  \bar \pro_1 &\bar \pro_2            \nn \\
                  \pro_1 &\pro_2
                 \end{array} \right)     ,
\ea
\nn
\eea
\bea
\ba{cc}
    \pro_ix^k=q^{3-2i}\delta_i^k+qY^{nk}_{mi}x^m\pro_n,  &
                        {\bar\pro}^i\bar y_j=\delta^i_j+qy_m\bar\pro^n\hat
                        R^{mi}_{nj},  \\
    \pro_i y_j=q(\hat R^{-1})^{lk}_{ji} y_k\pro_l,   &
                        \bar \pro^i x^j = \qin \hat R^{ij}_{kl}
                        x^k\bar\pro^l,
\ea                                        \label {comm}
\eea
\bea
    Y^{ri}_{sj} = {\rmm}^{ir}_{js} q^{2(s-i)}. \nn
\eea
The commutation relations (\ref {comm}) do
not differ on principle from
ones given in ref. [11]. Our choice is
motivated by using manifest covariant tensor notations which
 are convenient
in constructing explicit representations for the $q$-Lie algebras.
Thus, one implies all geometric objects with upper (lower)
indices to be transformed under the quantum group co-action
$ \Delta $ like classical co-(contra-) variant tensors.
For instance, a second rank tensor $N_i^j$
will be transformed as follows
\be
      (N_i^j)^{\prime} = (T^{\dag})^k_i {T^j}_l N_k^l \nn
\ee
(Hereafter the signs $ \bigotimes$ of tensor product are omitted).

      Let us define the left-invariant first-order differential operators
\bea
    D^{++}\equiv x_i\bar\pro^i, \ \ \ \ \
    &   D^{--}\equiv -y_i\pro^i,         \nn
\eea
where ($\pm \pm$) correspond  to \uone charges ($\pm 2$).
The action of the operators
  $ D^{\pm\pm}$ on the coordinates $(x, y)$ has a simple form
\bea
    D^{++}x^i=0,\ \ \ \ \   &    D^{--}x^i=y^i,          \nonumber\\
    D^{++}y_i=x_i, \ \ \ \ \  &  D^{--}y_i=0.         \nn
\eea
 The Leibnitz rule for these differential operators
 may be written in a  convenient form if one
considers their action  on functions with definite
\uone charges. The functions are defined in analogy with the
classical case
  [10] and can be decomposed in formal series
\be
   f^{(n{\geq}0)}(x,y)=\sum_{k=1}^{\infty}
C_{(i_1i_2...i_{k+n}j_1j_2...j_k)}
        x^{i_1}x^{i_2}...x^{i_{k+n}}y^{j_1}y^{j_2}...y^{j_k},      \nn
\ee
where $
C_{(i_1i_2...i_{k+n}j_1j_2...j_k)}$  are
   $á$-number coefficients symmetrized over all indices.
Functions with negative charges are defined in a simillar manner.
After some calculations one can find
the next Leibnitz rule for the operators $D^{\pm\pm}$:
\be
   D^{\pm\pm}(f^{(m)} g^{(n)}) = (D^{\pm\pm}f^{(m)}) g^{(n)}+
                         q^{-m}f^{(m)} D^{\pm\pm}g^{(n)}.
                                             \label {def2}
\ee

     This is a special feature of quantum group non-commutative geometry
that the quantum analogue to  classical
\uone generator
 can be realized as a second-order differential operator
\be
   D^0\equiv -x_i\pro^i-q^2 y_i\bar\pro^i +
                    (1-q^2)x_iy_k\bar\pro^k\pro^i.    \nn
\ee
The operator  $D^0$ has eigenfunctions which are just the functions with
definite \ \uone \  charges
\bea
   D^0f^{(n)} = {\{n\}}_q f^{(n)}, \ \ \ \ \ & \{ n\} _q \equiv
   \dfrac {1-q^{-2n}}    {1-q^{-2}} ,             \label {def3}
\eea
where    $ {\{n\}}_q $ \  is a $q$-number. It is not hard to check the
following Leibnitz rule for the operator
 $D^0$
\be
   D^0(f^{(m)}g^{(n)}) = (D^0f^{(m)})g^{(n)} + q^{-2m}f^{(m)}
                                     D^0g^{(n)}.           \label {def4}
\ee
Reducing the space of functions on \su to the space of functions
 with a definite  \uone charge one obtains
 the covariant description of the
coset $S_{q}^{2} \, \sim \,  {\su } / {\uone}$.

By direct calculating one can verify that the operators
$D^{\pm\pm, 0}$ form the $q$-deformed Lie
algebra of \su \ [12]
\bea
\ba{cc}
  {[D^0,D^{++}]}_{\dpst {q^{-4}}} = {\{ 2\}} _q D^{++}, \ \ \ &
   {[D^0, D^{--}]}_{\dpst {q^4}} = {\{ -2\}} _q D^{--},\\
   {[D^{++}, D^{--}]}_{\dpst {q^2}} = D^0,                     &
 \label{Dalg}   \\
\ea
\eea
here, $[A,B]_{\dpst {q^s}} \equiv AB-q^s BA$.
Note, that the algebra (\ref{Dalg}) is valid
irrespective of whether one imposes the unimodularity
constraint ${\cal D}=1$.
We shall treat the algebra (\ref{Dalg}) as a main $q$-deformed Lie algebra
of left-invariant vector fields on the quantum group \su.
A corresponding $q$-generalized Jacobi identity is available
\bea
&    [D^0, [D^{++}, D^{--}]_{\dpst {q^2}}]
 + [D^{++},[D^{--},D^0]_{\dpst {q^{-4}}}]_{\dpst {q^{-2}}} \nn \\
&            +q^2[D^{--},[D^0,D{++}]_{\dpst {q^{-4}}}]_{\dpst {q^{-2}}} = 0.\nn
\eea

Let us  now  pass to constructing other possible $q$-deformed
Lie algebras of
left-invariant vector fields on the \su. For this purpose we consider
 differential operators $\mu,\,\,\nu\,\,$ [13]
 \be
   \mu = 1+(q^2-1) y_i{\bar {\pro}}^i,  \,\,\,\,\,\,\,
   \nu = 1+ (1- \frac{1}{q^2}) x_i \pro^i.   \nn
\ee
One can see that the operators
 $\mu,\,\,\nu\,\,$ obey the simple commutation
relations
\bea
   \mu D^{--} = q^2 D^{--} \mu,  & \,\,\,\,\, \nn
   \mu D^{++} = \dfrac {1}{q^2} D^{++}\mu,                  \nn           \\
   \mu D^0 = D^0 \mu,            & \,\,\,\,\,  \mu\nu=\nu\mu  .   \nn
\eea
Similar formulae hold for the operator $\nu \,$ as well.
Using these relations one can find that
 the operators
    $\dpp,\,  \dmm ,\, \docal $ defined by the next equations
\bea
\ba{cc}
    \dpp = \mu^{-\frac {1}{2}} D^{++},& \dmm = \nu^{-\frac{1}{2}} D^{--} ,\\
   \docal = \qin \mu \nu D^0  \equiv {[\pro^0]}_q &
\ea                                             \nn
\eea
generate just the Drinfeld-Jimbo quantum algebra
\bea
 &  [\pro^0, \dpp ] = 2\dpp ,    &   [\pro^0, \dmm ] = -2\dmm , \nonumber  \\
 &  [\dpp ,\dmm ] = {[\pro^0]}_q.  &
\eea

 To construct  other possible $q$-Lie algebras
one introduces another  differential operators
$\Delta^{++}, \,\Delta^{--},\, \Delta^{0}  \,$ as follows
\bea
\ba{cc}
\Delta^{++}=D^{++}, & \Delta^{--}={\dfrac{q^2-1}{q^{2p}}}{\hat Z}^{1-p}D^{--},\\
\Delta^{0}=\dfrac{1-{\hat Z}^s}{1-q^{2s}},&\label {zet}  \\
\hat Z \equiv (\mu \nu)^{-\frac{1}{2}}, & \hat Zf^{(n)}=q^n f^{(n)} .
\ea
\eea
The operators $\Delta^{\pm\pm,0}$ generate
the next $q$-deformed Lie algebra:
\bea
\ba{cc}
\Delta^{++}\Delta^{--} - q^{2p}\Delta^{--}\Delta^{++} =
\dfrac{{\hat Z}^2-1}{{\hat Z}^{1+p}},  \label {quad}   \nn &         \\
\Delta^{0}\Delta^{++}-q^{2s}\Delta^{++}\Delta^{0}=\Delta^{++},\nn &    \\
\Delta^{0}\Delta^{--}-q^{-2s}\Delta^{--}\Delta^{0}=-q^{-2s} \Delta^{--}, &
\ea \nn
\eea
where $s,p$
    --  arbitrary integers. The equation (\ref {zet}) allows to express
the operator $\hat Z$ \  in terms of $\Delta^{0}$, then the
arbitrariness in the choice of
parameters $s, p$ can be reduced by considering only quadratic in
 $\Delta^{\pm\pm,0}$ $q$-commutators.

\section{ Gauge covariant differential algebra}
\setcounter{equation}{0}
%\stepcounter{section}
In this section we give description of possible  \su bicovariant
differential algebras with \uone conserved charge. The gauge covariance
condition leads to a unique differential algebra of \su. At the same time
a Leibnitz rule for the exterior differential is not fixed yet.
To find the differentiation rules one needs to choose a corresponding
 $q$-Lie algebra of left-invariant vector fields.

    Consider the left-invariant Cartan 1-forms
$\Omega$ on the quantum group \su
\bea
\Omega=dT^{\dag} T \equiv \left(\begin{array}{c cr}
        \oo &\op                         \\
        \om &-q^{2}\oo
                       \end{array}
                                  \right),         \nn
\eea
where $ \oo, \op, \om $ are the basic left-invariant differential 1-forms
with corresponding  \uone charges $( 0, +2, -2)$. One defines
gauge transformations as follows
\bea
    T^{g}=\tilde {T} T, \,\,\,\,\,\,\,\,\,\,\,\,\,\,\,\,\,\,\,\,\,\,\nn  \\
   \Omega^{g} = \Omega - T^{\dag} \tilde {\Omega}T,  \,\,\,\,\,\,\,\,
                                          \tilde{\Omega} \equiv d\tilde
                                        {T}^{-1}\tilde{T},
\eea
where the matrix $\tilde{T}$ commutes with the matrices
 $T, dT$ and satisfies the same equation
(\ref {mainrel}) as for the matrix $T$. In the case of
gauge symmetry the matrices
$\tilde{T} , T $ depend on the coordinates of a base space-time
and the connection 1-form $A$ has the same transformation and
commutation properties as the right invariant
1-form $dT T^{\dag}$.
It turns out that the requirement of
 global $U(1)$-covariance and
the consistence with the quantum group structure determine
uniquely all commutation relations
between the differential 1-forms $\omega$ and the coordinates $(x,y)$.
As a result we have
\bea
\ba{cc}
\op x=qx\op , &   \om x= \qin x\om+ \dfrac{1-q^{4}}{q} y\oo,     \\
\op y=\qin y\op, &  \om y=q y \om,\\
\oo x= x\oo + (1-\dfrac{1}{q^2}) y \op, & \oo y=y\oo .
\ea
\eea

Similar consideration of commutation relations for the basic
 differential 1-forms ${\omega}^{\pm\pm,0}$ leads to covariant
algebras parametrized by a real number $\sigma$:
\bea
\ba{cc}
\op \op =\om \om = 0,  &\omega^{\pm\pm} \oo + q^{\pm 2} \oo
\omega^{\pm\pm} =0, \label{f4alg}
\ea
\eea
\bea
\op \om + q^{\sigma} \om \op +\dfrac {q^2 (1-q^{\sigma})(1+q^2)}{q^2-1}
                       \oo \oo =0,  \label{f4alg4}
\eea
\bea
\oo \oo = \dfrac{1-q^2}{q^2 (1+q^2)} \op \om.       \label {f4alg5}
\eea

It should be noted that the algebra defined by eqs.
(\ref{f4alg}-\ref{f4alg5})
 is bicovariant irrespective of  whether one considers
  the last relation (\ref {f4alg5}).
  Requiring the covariance under the gauge transformations and
  using the additional
  commutation constraint
  \bea
\tilde {\Omega} \Omega =-q^2 \Omega \tilde {\Omega}    \nn
\eea
   one finds a unique gauge covariant differential algebra
at   $\sigma =4$:
  \bea
\ba{cc}
\op \op =\om \om = 0,  &\omega^{\pm\pm} \oo + q^{\pm 2} \oo
\omega^{\pm\pm} =0, \\
(1+q^2)^2 \oo\oo =
 \dfrac{1}{q^2}\op\om+q^2 \om\op.    &
\ea
\eea

The equation (\ref {f4alg5}) is not gauge covariant and should be omitted.
 So defined gauge covariant differential
algebra differs from one considered in refs. [6,9].Our treatment
does not contain the condition of vanishing for the central element
$     C_2 \equiv tr_{q} (\Omega^2),$
which is not gauge covariant. Here we have used the
 notion of the $q$-deformed covariant
trace  [4, 7].

 One can rewrite the commutation relations for the gauge
 covariant differential algebra
in terms of the $R$-matrix. Direct checking leads to
the next formulae
\bea
%\ba{cc}
& R_{12} dT_{1} T_{2} = T_{2} dT_{1}  R_{12},    \label {f4algdif}
% &
     \\
& R_{12} \Omega_{2} R^{-1}_{12} \Omega _1 + \dfrac{1}{q^2}
    \Omega_1 R_{12} \Omega_2 R^{-1}_{12} 
  -    \dfrac {q^2}{1+q^2 +q^4} (E_{12} - (1+q^2)A_{21}) tr_q
                \Omega^2 = 0 ,
% &
%\ea
\eea
here $E^{ij}_{kl}=\delta^i_k\delta^j_l$
and $A_{21}$ is the quantum antisymmetrizer [11]. Note, that
the first relation
in (\ref{f4algdif}) had been obtained earlier in ref. [9].

To construct an exterior differential it is convenient to use the
 definition  based on the dualism between the
exterior algebra of differential forms and the $q$-Lie algebra
of vector fields. In this way
the Leibnitz rule is followed straightforwardly and
it depends only on a special choice  of the $q$-Lie algebra.

Let us start from a general 3D $q$-Lie
 algebra of left-invariant vector fields
    $D^a = (D^{++}, D^{--}, D^{0})$ on the quantum group \su with
    a Lie bracket
    \bea
[D^a , D^b]_{B} \equiv D^a D^b - B^{abcd} D^c D^d   = C^{abc} D^c .\nn
\eea
We consider the matrix $ B^{abcd} $ to be unitary,
so that it generates a representation of the permutation group.
Thus, one can easily define the alteration rules for the
tensor algebra of vector fields. Moreover, a generalized Jacobi
identity will be available as well.

The basic left-invariant differential 1-forms $\omega^a$ are defined
 as dual
 objects by means of the scalar product
$\omega^a (D^b) = \delta ^{ab}\;.$
The action of the exterior differential on arbitrary functions $f$ and
differential 1-forms $u$ is defined in analogy with the
 classical case [14]
\bea
\ba{cc}
df(D^a) = D^a f, &  \\
du(D^a, D^b) = -\dfrac{1}{2} (D^a u(D^b) - B^{abcd} D^c u (D^d) -
       u([D^a, D^b]_{B}),& \label {f5rules}    \\
                         du(D^a, D^b) = - B^{abcd} d u(D^c, D^d).&
\ea
\eea
Rules  for the exterior differentiation
 of the differential ($n>1$)-forms can be
generalized in a similar fashion.
 The Cartan-Maurer equations have a standard
form
\bea
d\omega ^d (D^a, D^b) = \dfrac{1}{2} C^{abc} \omega^d (D^c). \label {cm}
\eea
    As a concrete example we consider
 the $q$-Lie algebra
(\ref{Dalg}) which is consistent with the gauge covariant algebra of
left-invariant differential 1-forms (\ref{f4algdif}). In that case
differentiation rules (\ref{f5rules}) can be rewritten
 in a more familiar form after using the
 explicit tensor representation for the exterior products
of 2-forms $\omega^a \wedge  \omega^b$. After some calculations one finds
\bea
\ba{cc}
   df=\omega^a D^{a}f,   &      \\
   d(\op f)= d\op f +\beta \oo \op D^0 f -{d\omega^0} D^{++} f,   &  \\
   d(\om f) = d\om f +\beta q^2 \oo \om D^0 f + q^2 d\omega^0 D^{--}f, & \\
   d(\oo f)=d\omega^0 f+
 \beta q^2 \op \oo D^{--}f +\beta \om \oo D^{++}f,&\\
    \beta \equiv \dfrac{1+q^4}{q^2(1+q^2)}\,\,.    &
\ea
\label {f5rules2}
\eea
   It should be noted that the formulae (\ref{f5rules2}) involve just three
   independent basis differential 2-forms
 $\oo \op,  \oo \om , d\omega^0$
    in the space of exterior 2-forms in correspondence with
the classical case.
The fourth linearly independent basis 2-form $\sigma^0$ can be defined as
follows
\bea
 \sigma^0 =  \dfrac{1}{1+q^2}(\op\om+q^2\om\op),     \nn
\eea
The form $\sigma^0$ takes a non-zero value only for the
symmetrical tensor product $D^0\bigotimes D^0$:
\bea
\sigma^0(D^0,D^0) = \rho, \nn
\eea
where the number $\rho$ vanishes in the classical
 limit $q\rightarrow1$. Due to this
property the form $\sigma^0$ does not appear in eqs. (\ref{f5rules2}).

   To construct the differentiation rules for the
 ($n>1$)-forms it is convenient to use the specific structure of the
exterior algebra of \su. It is easy to check that the invariant 2-form
\be
C \equiv {\dfrac{\beta q^4}{1+q^2+q^4}} C_2 = {\dfrac {\beta q^2}{1+q^2}} (\op\om
+ \om\op)  \nn
\ee
is a central element. We put the natural constraint
\be
d(C{\omega}^{(n)}f) = Cd({\omega}^{(n)}f). \nn
\ee
Using this constraint and starting from the most general
form for differention rules one finds
 \bea
     \ba{cc}
    d(\oo\omega^{\pm\pm} f)=\pm q^{\mp 4} {\upsilon} D^{\pm\pm}f, &  \\
        d(\op\om f)=\dfrac{1}{\beta} C df + {\upsilon} D^0 f, & \\
    d(\om\op f)=\dfrac{1}{\beta q^2} C df - {\upsilon} D^0 f, &
                                         \label {drules2} \\
    d(\oo\op\om f)=\dfrac{1}{\beta}C d\omega^0 f-q^4 C\oo\op D^{--}f
                   -\dfrac{1}{q^2} C\oo\om D^{++}f, &\\
    d({\upsilon}f)=0,&   \\
 \upsilon \equiv \dfrac{1}{2}
                         (\oo\op\om-q^2\oo\om\op)\,,&
     \ea
 \eea
where $ \upsilon$ is a volume 3-form on the \su.

All basis differential forms of order
$n>3$ can be obtained from lower order forms multiplied by the
invariant $C$ in an appropriate degree. So that the relations
                        ( \ref {f5rules2}, \ref {drules2})
complete the differential rules for the differential algebra of \su.

    Having carried out some calculations one can also find the
explicit expressions for
    the Cartan-Maurer equations (\ref {cm})
    \bea
    d\Omega=\Omega^2-\frac{q^2}{1+q^2}{ I }\, tr_q \Omega^2. \nn
           \eea
The right hand side of the equation contains only the
traceless part of
 $\Omega^2$.

\section{Construction of the minimal $SU_q(2)$  gauge theory }
\setcounter{equation}{0}
%\stepcounter{section}

A minimal gauge theory corresponding to the quantum algebra \su
was proposed in ref.
[15], where the initial gauge transformations contain
by definition the antipodal map.
One can try to formulate a gauge Yang-Mills theory for the quantum group
\su in analogy with a covariant $GL_q(N)$ version proposed
in ref. [7]. For this purpose one should define
the algebra of main operators (matter fields and gauge potential)
and a corresponding comodule structure.

 In the case of the quantum group \su it is convenient to put
the matter field $\phi^i (\bar{\phi^i})$ into one matrix $\Phi =
(\bar{\phi^i} \ \phi^i)$ .
   The operators $(\Phi, d\Phi)$ generate the $Z_2$-graded algebra $Z$
with the same commutation relations as for the differential algebra
of \su. The algebra $Z$ is a left \su-comodule with the following co-action:
\be
  \Phi \rightarrow \Phi^g = T\Phi,
\ee
\be
  d\Phi \rightarrow (d\Phi)^g = (dT)\Phi +T(d\Phi).
\ee
All axioms for the comodule are fulfilled.

We introduce also an operator $A$ (gauge potential 1-form) satisfying
the same commutation relations as for the right-invariant Cartan 1-forms
on the \su. One can consider the quantum analogue to
the gauge transformation for
the $A$:
\be
 A \rightarrow A^g=TAT^{\dag} + (dT)T^{\dag}.
\ee
A covariant differential $\Delta$ acting on the matter field is defined as
\be
\Delta \Phi =(d-A)\Phi.
\ee
The curvature 2-form $F$ is introduced as follows
\be
F=dA-A^2+\dfrac {1}{1+q^2}tr_qA^2
\ee
and it contains only traceless part.

   To define the commutation relations in the algebra $G$ generated by
the operators $(\Phi, d\Phi, A, dA)$ we will take into account the
compatibility conditions with the gauge transformations and the centrality
property for the quantum determinant ${\det}_q\Phi =1$. The commutation relations
for the operators $\Phi, A$ are uniquely defined from the
formula (\ref{f4algdif}), since the gauge potential
$A$
in a pure gauge limit
is just the right-invariant Cartan form on the quantum group \su,
so we have
\be
\Phi_1 A_2=R_{21}A_2 R^{-1}_{21}\Phi_1.
\ee
We demand the covariant combination ($F\Phi$) to have the same commutation
relations with itself and $A$ as for the operator $\Phi$
\bea
(F\Phi)_1 A_2=R_{21}A_2 R^{-1}_{21}(F\Phi)_1,            \\
R_{12}   (F\Phi)_1 (F\Phi)_2=(F\Phi)_2 (F\Phi)_1 R_{12} . \label{11}
\eea
Using these formulae one can easily derive the next commutation relations:
\bea
F_1R_{21}A_2 R^{-1}_{21}=R_{21}A_2R^{-1}_{21}F_1,
F_1R_{21}{\tilde{r}}_2 R^{-1}_{21}=R_{21}{\tilde{r}}_2R^{-1}_{21}F_1,
\eea
where $
\tilde{r}=(d\Phi){\Phi}^{\dag}$ .

The bicovariance condition, the traceless property for the $F$ and
the centrality of the quantum determinant ${\det}_q \Phi$ lead
to the following commutation relation for the operators $(\Phi, F)$:
\be
\Phi_1 F_2=R_{21}F_2 R^{-1}_{21}\Phi_1.
\ee
From this equation taking into account the eqn. (34) one finds
\be
R_{12}F_1R_{21}F_2=F_2R_{12}F_1R_{21}.
\ee
Note, that the last equation coincides with the corresponding relation
in $GL_q(N)$ gauge theory [7]. It is not difficult to check the next
simple relations as well:
\bea
R_{12}\Delta \Phi_1 \Phi_2=\Phi_2 \Delta \Phi_1 R_{12},\\
\Phi_1 dA_2 = R_{21}dA_2 R^{-1}_{21}\Phi_1.
\eea
The commutation relations obtained above are the main ones which imply
all other commutations in the algebra $G$.
     To construct a formal expression for the Lagrangian of the gauge
theory one needs to specify the underlying space-time and define the
dual $\ast$-operation. The simplest variant correspond to the choice
of the space-time isomorphic to the quantum space of \su.
Another problem is the construction of Leibnitz rules for the exterior
differential which have essentially non-standard form.
Differentiation rules for the algebra of matter
fields $(\Phi^i, d\Phi)$
are defined by similar formulae (23, 26) in full analogy with the case of the differential
algebra of \su.
 For instance, one has the following equations
 for the quadratic combinations of matter fields
 \bea
\ba{cc}
 d(\phi^i \phi^j) = d\phi^i \phi^j +\dfrac{1}{q^2} \phi^i  d\phi^j +
       \dfrac{q^2-1}{q^3} \vareps^{ij} d\phi_{k} \phi^{k}, &      \\
 d(\bar {\phi}^i \bar {\phi}^j) = d\pbar^i \pbar^j +q^2 \pbar^i d\pbar^j,
                                    &   \\
 d(\pbar^i \phi^j) = d\pbar^i \phi^j + q^2 \pbar^i d\phi^j +
  (q^2-1)\pbar^i \pbar^j d\phi_k\phi^k.   &
\ea  \nn
  \eea
     Another possible way towards a consistent minimal quantum
group gauge Yang-Mills theory
   corresponds to the differential calculus with a $q$-Lie algebra
   differed from one defined by eqs. (\ref{Dalg}).\\
{\bf Acknowledgments}
$$~$$
The author would like to acknowledge Professor M. Arik and the
members of the Organizing Commitee for kind hospitality and financial support.
Author thanks F. Mueller-Hoissen, V.D. Gershun and M. Nomura for
useful discussions.
$$~$$

\end{document}